\pdfoutput=1
\documentclass{revtex4}
\usepackage{latexsym}
\usepackage{amsfonts}
\usepackage[latin1]{inputenc}
\usepackage[caption=false]{subfig}
\usepackage{graphicx}
\usepackage{mathrsfs}
\usepackage{amssymb}
\usepackage{amsmath}
\usepackage{color}
\usepackage{fancyhdr}
\newcommand{\gtap}{\stackrel{\displaystyle >}{\,_{\! \,_{\displaystyle
\sim}}}}  

\usepackage{color}

\begin{document}
\begin{flushright}
SHEP-13-11\\
PSI-PR-13-08
\end{flushright}

\title{Production of $Z^\prime$ and $W^\prime$ via Drell-Yan processes in the \break 4D Composite Higgs
Model at the LHC}

%

\author{D. Barducci, A. Belyaev, S. Moretti}
\affiliation{School of Physics and Astronomy, University of Southampton, Highfield, SO17 1BJ, UK.}
\email{d.barducci,a.belyaev,s.moretti@soton.ac.uk}
\author{S. De Curtis}
\affiliation{INFN, Sezione di Firenze,
Via G. Sansone 1, 50019 Sesto Fiorentino, Italy.}
\email{decurtis@fi.infn.it}
\author{G.M. Pruna}
\affiliation{Paul Scherrer Institute, CH-5232 Villigen PSI, Switzerland.}
\email{giovanni-marco.pruna@psi.ch}

\begin{abstract}
We present an analysis of both the Neutral Current (NC) and Charged Current (CC)
Drell-Yan processes at the LHC within a 4 Dimensional realization of a Composite Higgs
model studying the cross sections and taking
into account the possible impact of the extra fermions present in the spectrum.
\end{abstract}

\maketitle

\thispagestyle{fancy}


\section{Introduction and model description}
Composite Higgs models with the Higgs boson arising as a pseudo-Goldstone state provide an elegant solution to the hierarchy
problem present in the Standard Model (SM) suggesting an alternative pattern leading to the mechanism of spontaneous
electroweak symmetry breaking. The general idea proposed in \cite{Georgi:1994ha} and based on the 
minimal coset $SO(5)/SO(4)$ in \cite{Agashe:2004rs}
has been specialized into a 4 Dimensional (4D) description called the 4D Composite Higgs Model (4DCHM) in \cite{DeCurtis:2011yx}.

\begin{figure}[h!]
\includegraphics[width=7.5cm]{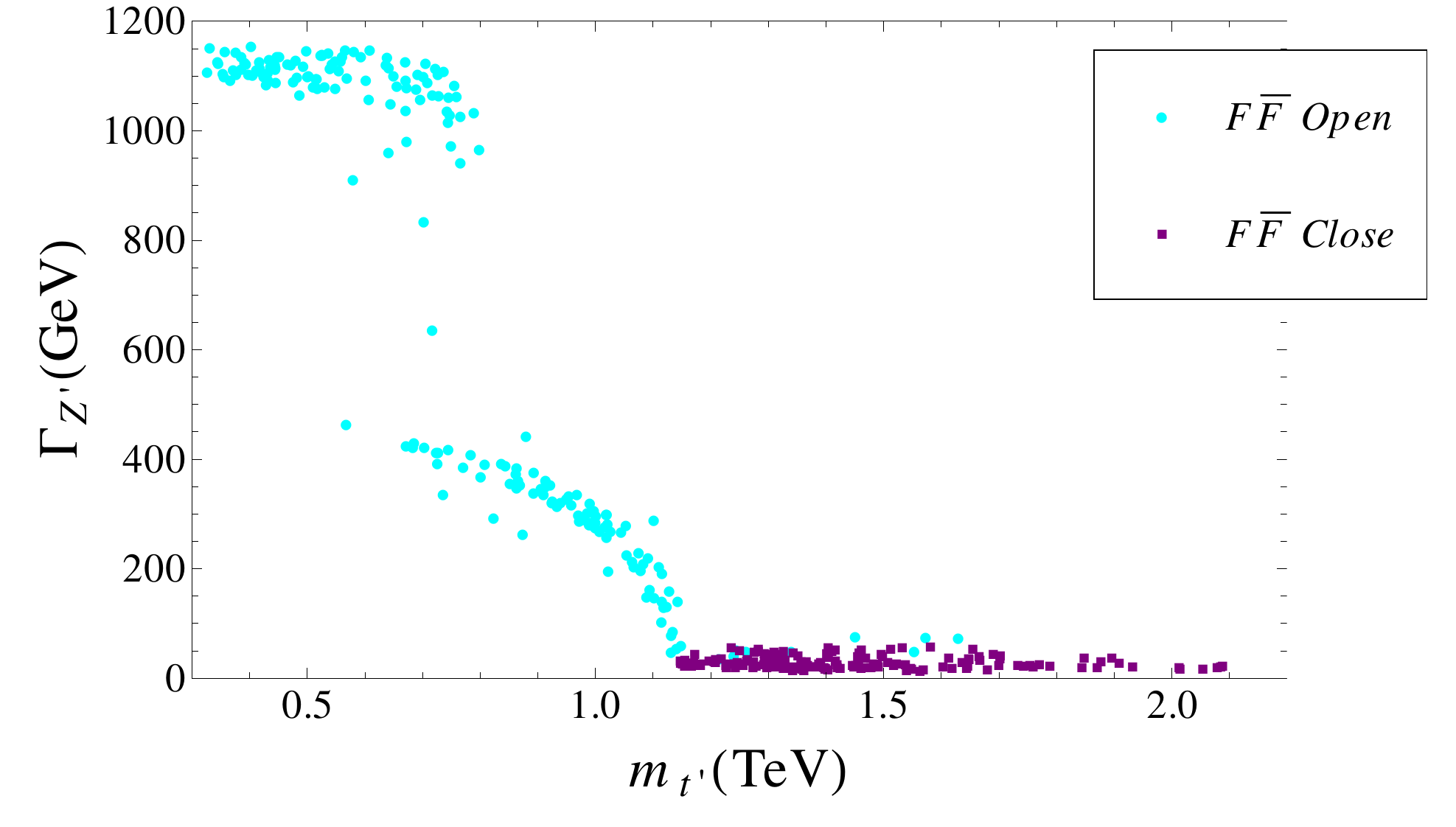}
\includegraphics[width=7.5cm]{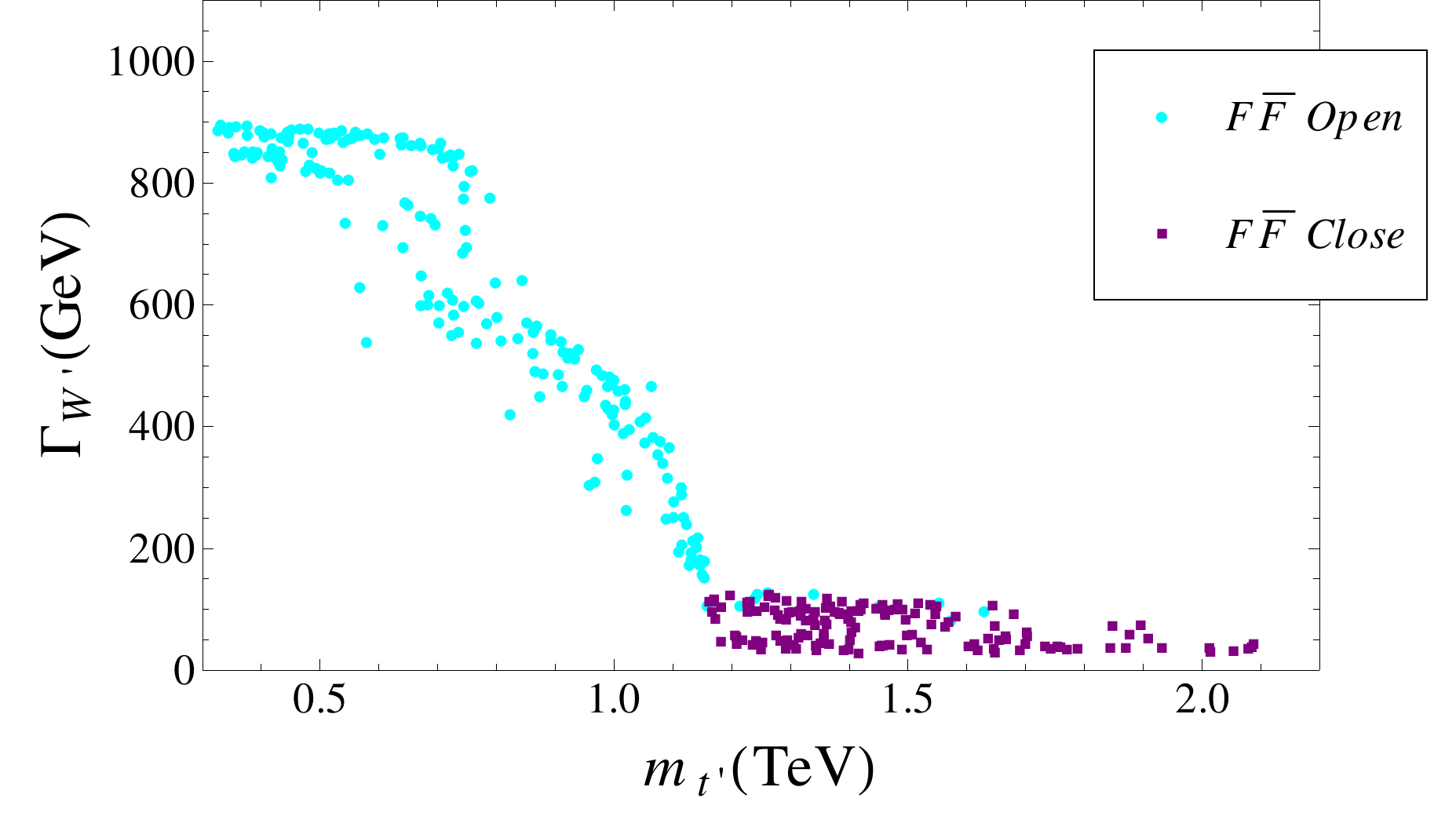}
\caption{Width of the $Z^\prime$ and $W^\prime$ in function of the mass of the lightest $t^\prime$. In purple points
where the decay in a pair of heavy fermions $F\bar F$ is kinematically not allowed, in cyan the points where this processes is allowed. }
\label{fig:width}
\end{figure}

Besides the SM content the 4DCHM spectrum consists of 5 $Z^\prime$ and 3 $W^\prime$ with masses and couplings
described by two parameters, the compositeness scale $f\gg v \simeq 246$ GeV and the coupling constant $g_{\rho}$
of the extra gauge fields, $4 \pi \gg g_{\rho} \gg g$, where $g$ is the SM gauge coupling.
Heavy fermions with both SM, herein collectively called
$t^\prime,b^\prime$, and exotic electric charges are also present and the value of
their mass strongly modify the widths of extra gauge bosons as shown in Fig. \ref{fig:width}.

\section{Drell-Yan analysis}
\begin{figure}[h!]
\includegraphics[width=4.2cm,angle=90]{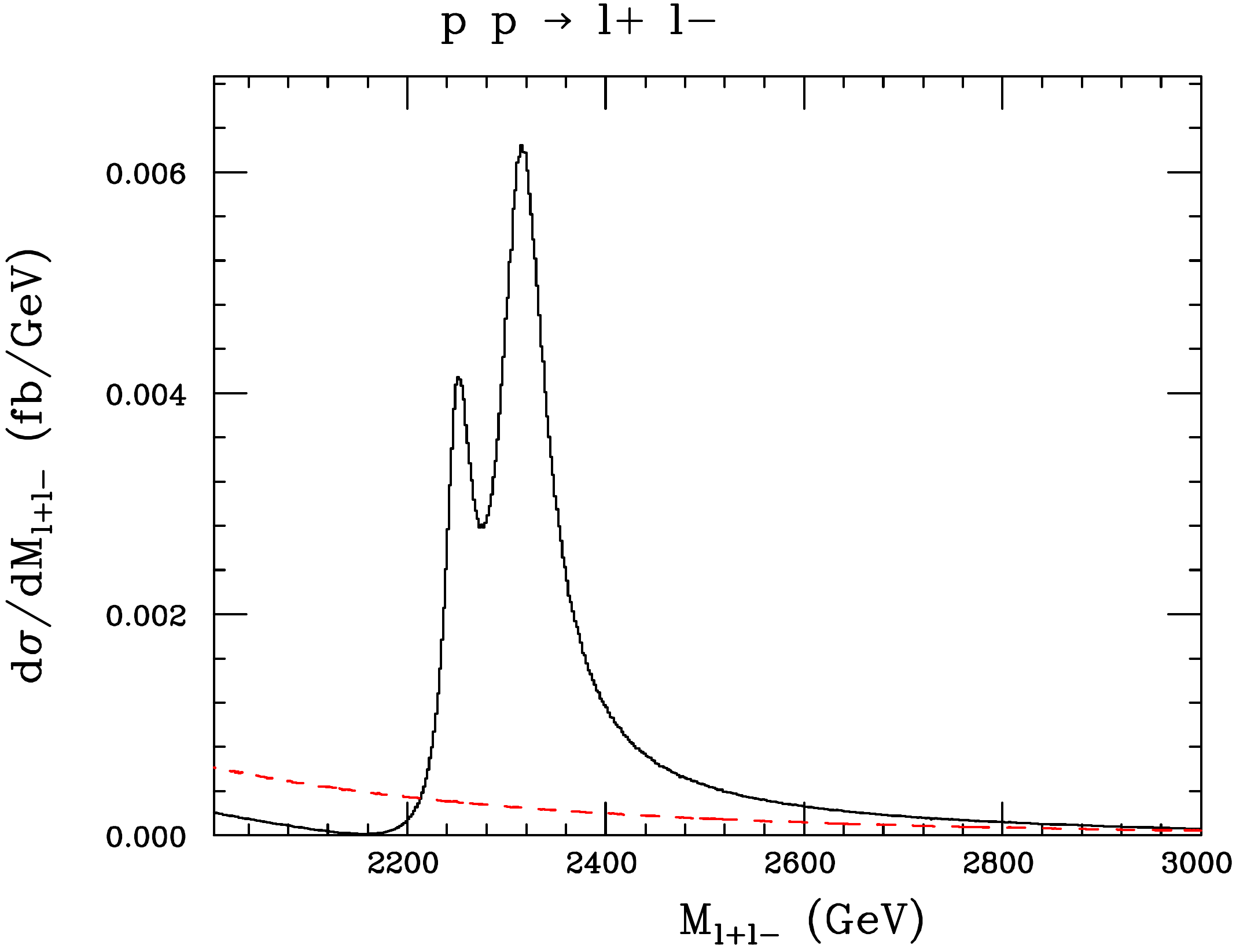}
\includegraphics[width=4.2cm,angle=90]{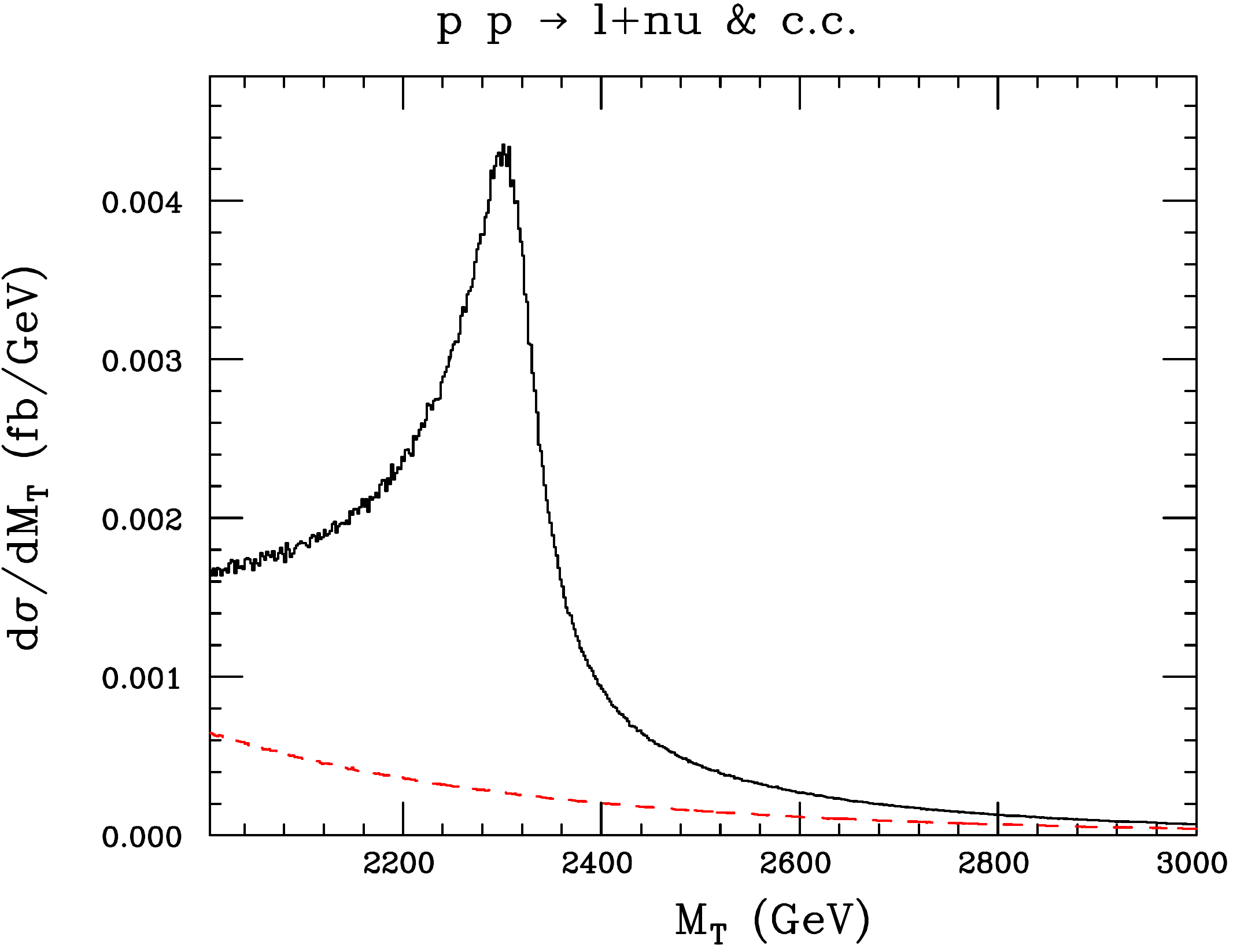}
\includegraphics[width=4.2cm,angle=90]{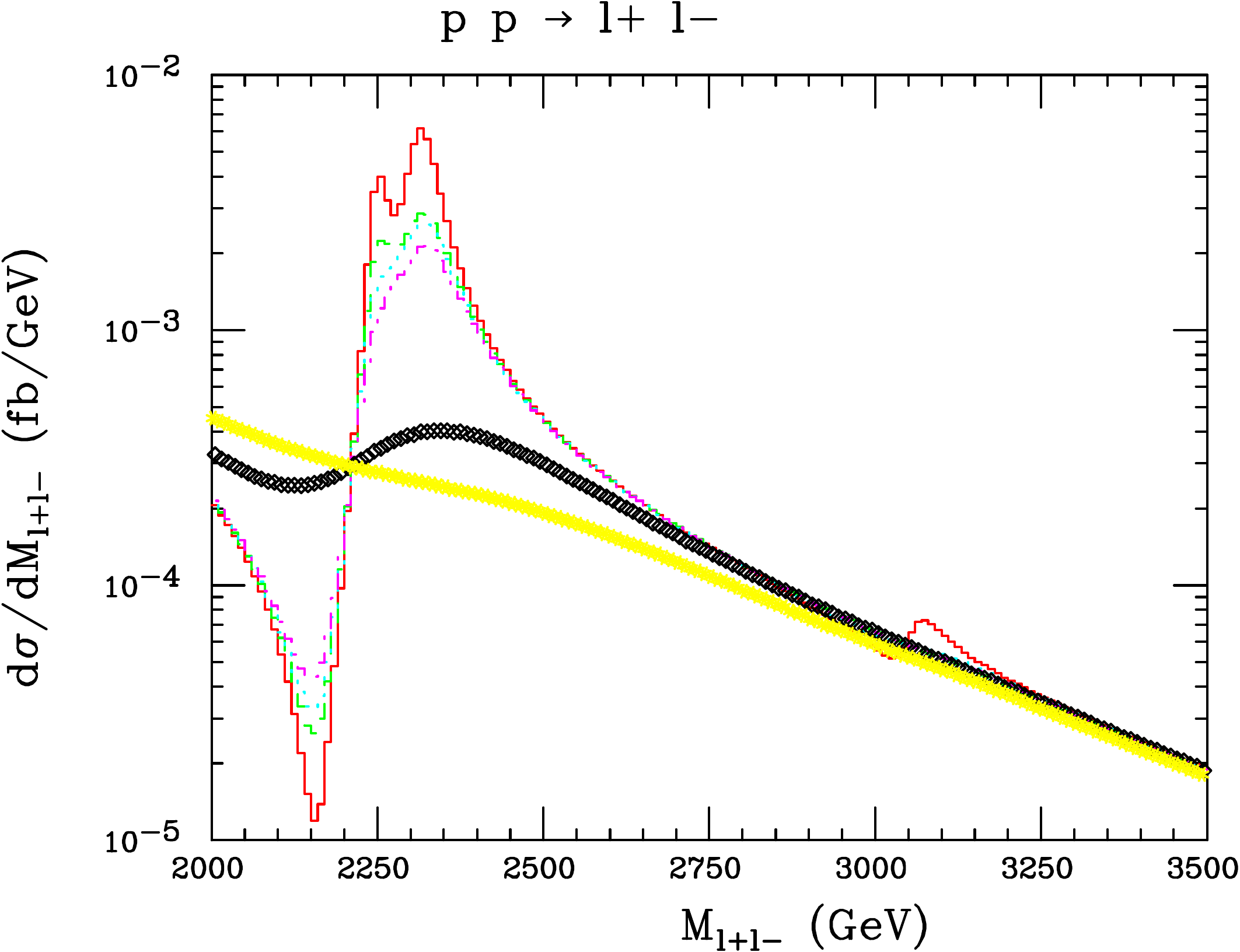}
\caption{Differential cross section of the NC (left) and CC (center) channels for $f=1.2$ TeV and $g_{\rho}=1.8$.
Integrated cross sections for the 4DCHM[SM], given as solid/black[dashed/red] lines, are 1.37[0.21] fb (NC)
and 1.33[0.23] fb (CC).
Line shape for the NC process (right) for the same combination of $f$ and $g_{\rho}$
for different values of the masses of the lightest $t^\prime$ (See Ref. \cite{Barducci:2012kk} for the meaning
of colors in the last plot). Results are for the 14 TeV LHC setup.}
\label{fig:crosssections}
\end{figure}

The Drell-Yan processes $ p~p \rightarrow l^+~l^-$ (NC) and $p~p \rightarrow l \nu_l+c.c.$ (CC)
for the 14 TeV stage of the Large Hadron Collider (LHC) run for the 4DCHM \cite{Barducci:2012kk,Barducci:2012as} have been studied
via the implementation of the model in automatic tools
(LanHEP \cite{Semenov:2010qt} and CalcHEP \cite{Belyaev:2012qa}) to perform a fast phenomenological analysis.
Constraints on the parameter space due to
electroweak precision tests have been taken into account
by requiring \break $m_{\rho}\simeq m_{Z^\prime,W^\prime}\gtap2\text{ TeV}$. Results for both the NC and CC channels in a regime where
the widths of the $Z^\prime$s and $W^\prime$s are small are shown in Fig. \ref{fig:crosssections}.

We have shown that the 14 TeV stage of the LHC places us in the position of studying the rich phenomenology of the gauge sector for
the 4DCHM that is so testable at the LHC for 2-3 TeV $Z^\prime$ and $W^\prime$, allowing, in certain regions of the parameter space, 
to distinguish the two lighter neutral resonances.
The results are obtained for the 14 TeV LHC but  remain essentially valid for the next LHC run at 13 TeV.

Conversely the possible presence of extra fermions with a mass lower than the TeV scale represents the first possibility
to test the 4DCHM already with the 7 and 8 TeV LHC data.

\begin{acknowledgments}
Some of us thank the NExT Institute for partial support.
\end{acknowledgments}

\bibliography{proceed_cagliari}

\end{document}